# AuraSight: Generating Realistic Social Media Data

Lynnette Hui Xian Ng, Bianca N. Y. Kang, Kathleen M. Carley
July 14, 2025
CMU-S3D-25-109

Software and Societal Systems Department
School of Computer Science
Carnegie Mellon University
Pittsburgh, PA 15213

This work was supported in part by the following grants: MURI: Persuasion, Identity & Morality in Social-Cyber Environments (N000142112749, Office of Naval Research)); Community Assessment (N000142412568, Office of Naval Research); Threat Assessment Techniques for Digital Data (N000142412414, Office of Naval Research) and the Knight Foundation. Additional support was provided by the center for Computational Analysis of Social and Organizational Systems (CASOS) and the Center for Informed Democracy & Social-cybersecurity (IDeaS) at Carnegie Mellon University. The views and conclusions contained in this document are those of the authors and should not be interpreted as representing the official policies, either expressed or implied, of the Office of Naval Research, the Knight Foundation, or the U.S. government.

Center for the Computational Analysis of Social and Organizational Systems
CASOS technical report.

**Abstract**

This document details the narrative and technical design behind the process of generating a quasi-realistic set X data for a fictional multi-day pop culture episode (AuraSight). Social media post simulation is essential towards creating realistic training scenarios for understanding emergent network behavior that formed from known sets of agents. Our social media post generation pipeline uses the AESOP-SynSM engine, which employs a hybrid approach of agent-based and generative artificial intelligence techniques. We explicate choices in scenario setup and summarize the fictional groups involved, before moving on to the operationalization of these actors and their interactions within the SynSM engine. We also briefly illustrate some outputs generated and discuss the utility of such simulated data and potential future improvements.

# Table of Contents

# 1 Introduction

Research on social media often involves observational studies of online user interactions and discourse. Predictive studies - where researchers simulate network dynamics to understand processes that lead to complex emergent behaviors and/or to perform controlled experiments on them - are less common. Nonetheless, they are just as important, as these simulation studies enable the exploration of questions that otherwise require unethical and unfeasible perturbation of behaviors in an actual social media environment. Some phenomena that can be studied with synthetic social media data include: opinion dynamics (Ng & Carley, 2022), polarization (Lu & Lee, 2024), and information propagation (Gurung et al., 2025). The customizability of synthetic data grants great potential in using them to study a range of very specific scenarios.

Good synthetic data should display several features:

1. Fidelity - Good synthetic data should accurately replicate the statistical properties, distributions, and structural characteristics of real-life data (Alaa et al., 2022; Yuan et al., 2024). Network topology, degree distributions, clustering coefficients and temporal dynamics that are fundamental to the social media platform should also be preserved.

2. Diversity - Good synthetic data should capture the full range of variations in the original data. This includes having a large spectrum of agent types and behaviors that are realistic to social media patterns (Chang et al., 2024). Variations in the data should also follow some logical consistency.

3. Stability - Good synthetic data should be stable. There should be consistent results, despite parameter variations, noise and minor perturbations. The data should be reproducible; equivalent datasets should be generated when the same parameters and procedures are applied (El Emam et al., 2024; Grund et al., 2024).

One technique to construct synthetic data draws upon network science. Synthetic network structures are created by applying well-established mathematical frameworks, resulting in realistic network topologies observable in real-world social media platforms. There are three key models. The first is the Erdos-Renyi model, the foundational random graph model (Erdos & Renyi, 1960). In the Erdos-Renyi model, nodes are connected following an independently and identically distributed probability distribution. The networks produced have a Poisson-type degree distribution. The second type is the Small-World network model. Initialized with a regular lattice, edges are probabilistically rewired to achieve the high clustering and short path lengths characteristic of a social network (Watts & Strogatz, 1998). The last key model is the Scale-Free network structure. Constructed through preferential attachment mechanisms, new nodes connect to existing nodes with a probability proportional to their degree. The resulting degree distribution is thus a power-law, mirroring hub-dominated structures of real social networks. For all three models, the resulting generated network structures form the topologies in which simulated agents are embedded in. Each model has its own unique topological constraints that differently capture the fundamental structural properties of social media networks (Albert & Barabási, 2002). Simulation studies thus have used all three in conjunction; for example, all three network structures were used to investigate the relation between confederate stance perturbation and the eventual overall stance of the full social network (Carragher et al., 2023).

Large Language Models (LLMs) offer a new lease on what is possible with generating synthetic networks. Increasingly, researchers are using LLMs for generating social network data, offering new paradigms for understanding complex social systems (Eberlen et al., 2017). For example, the language generation and understanding ability of LLMs can simulate the trends in public opinion and engagement of news across millions of agents (Zhang et al., 2025). The emotions, attitudes, and interactions of agents have also been modeled using LLM-driven social networks, with encouraging accuracy in capturing information and emotion spread (Gao et al., 2025).

However, these developments do not mean that techniques involving traditionally mathematically-defined networks are obsolete. While LLM-generated data does overcome some limitations of traditional mathematically-defined networks - like the limited ability to generate realistic social media content - they have trappings around the simulation of actual social interactions. LLM-constructed networks generally over-represent the principle of homophily, with clustering and echo-chambers often observed (Ferraro et al., 2025). To address this, a hybrid approach is sometimes used. Zhang et al. (2025) presented LLM-AIDSim, a model which integrated LLMs into an influence diffusion model. The use of the LLM (LLaMa 3) was centered around generating user agent profiles and full-text responses.

Following the lead of such work, work here at the CASOS center has also adopted a hybrid approach for our social media post generation pipeline (Hicks, 2024). The pipeline, AESOP-SynSM, combines past work in ABM and LLM simulations by (1) using the LLM to generate content and (2) relying on network science principles to specify underlying social networks to structure interactions. What AESOP-SynSM introduces is further customizability, flexibility, and more specific agent types. The pipeline is still under active development, but in brief, features include:

1. Accessibility for a non-technical scenario designer to construct a fictional scenario easily from scratch via the AESOP Graphical User Interface (GUI).
2. Research-informed agent types beyond simple 'human' *users* - for example, organizations and various bot types.

3. Generation of realistic tweet output (including replies, retweets, and quotes) following the user-defined AESOP scenario by SynSM, output in the Twitter API V1 JSON format.

We strongly encourage interested readers to refer to Hicks (2024) for the full background on initial builds of AESOP-SynSM. In this report, we present an end-to-end usage of the pipeline to generate a multi-day pop culture episode that happens on X. We also introduce some features (multiple bot types and a new class of actors) that were not available in the initial builds.

## 1.1   Brief Overview of The AESOP-SynSM Simulation Modeling Pipeline

The pipeline consists of two large components. First, AESOP - AI-Enabled Scenario Orchestration and Planning Tool - is a PySide6 GUI for easy scenario construction. Here, key actors, their communities (groups), events in the scenario and relevant narratives are specified. Next, SynSM - Synthetic Generation for Social Media - is the scenario generation engine. SynSM currently supports X and Telegram-type output.

The AESOP-SynSM pipeline thus starts by an end-user constructing a scenario on the AESOP GUI, before moving on to running the simulation on SynSM. Note that posts and interactions on SynSM are informed by AESOP via a set of JSONs that are generated from AESOP, following the end-user's specifications. SynSM generates social media content and outputs all interactions in JSONs following standard social media API definitions (SynSM). For X, we follow the Twitter API V1 JSON format (https://developer.x.com/en/docs/x-api/v1). Having the output in this format allows for straightforward applications of social media analysis and integration into visualization software like ORA-Pro. Figure 1 illustrates the flow of data across the pipelines.

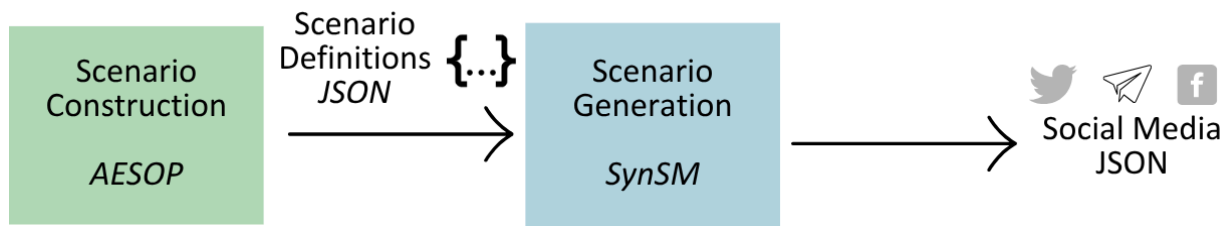

Figure 1: Illustration of pipeline for AuraSight scenario construction and generation.

Our fictional scenario, AuraSight, which we will introduce now, is designed to happen on X and we thus focus only on that platform in this report.

# 2 AESOP: The AuraSight Scenario

## 2.1 Broad Setting

AuraSight takes direct inspiration from the highly popular Eurovision Song Contest (ESC), a long-running annual song contest held by the European Broadcasting Union since 1956. While the contest has claimed to be non-political (*About the Eurovision Song Contest*, 2025), the constant presence of cultural and identity contests within it show otherwise (e.g., Dubin et al., 2022; Press-Barnathan & Lutz, 2020).

In our scenario, AuraSight is also a long-running song contest held within a fictional Northern Region. The organizers of AuraSight as well do not set out to politicize the event, but find extraneous conflicts often bearing down on the annual contest. Figure 2 shows the fictitious map of the Northern Region.

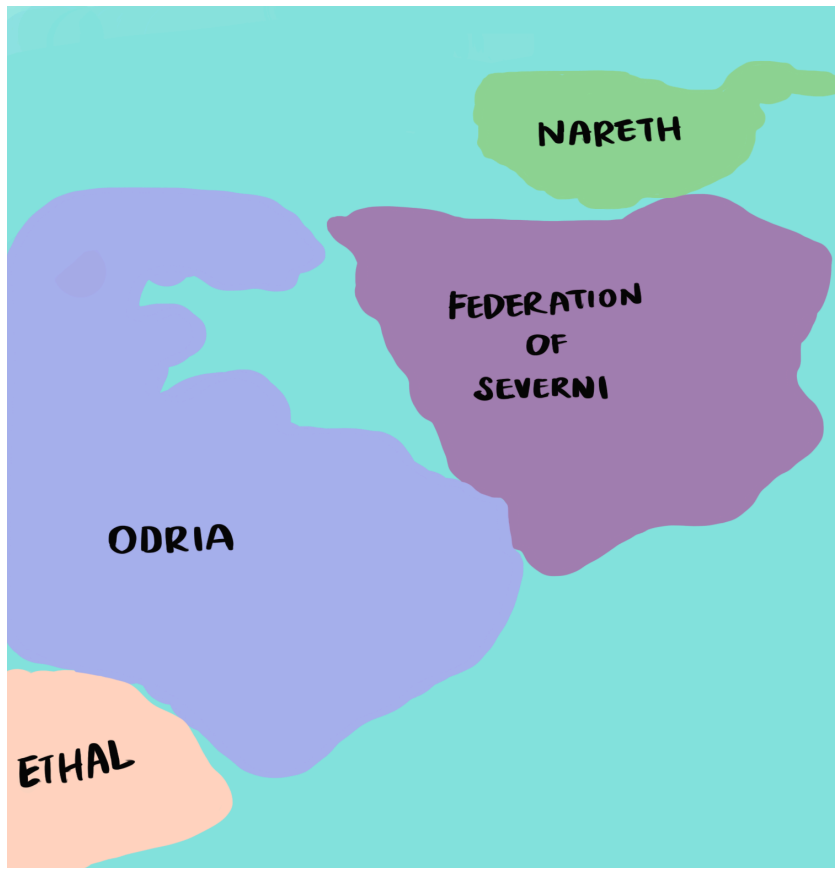

Figure 2: Map of Northern Region of AuraSight world.

We base the fictional conflict in our scenario on:

1. Alekseev's run to be Belarus' representative in ESC 2018; especially the discussions around (1) his switch from running for Ukraine to Belarus (Ko, 2018) and (2) whether singers should be nationals of the country they are representing (Ten Veen, 2018).

2. More loosely, the post-Soviet pop market, where countries (e.g., Russia, Ukraine, Belarus) often engage with the pop markets of their neighbors, though shifts have been observed after the 2022 invasion of Ukraine (Lee et al., 2024).

Our fictional scenario covers 3 days, focusing on one country's representative selection process leading up to AuraSight 2030, which will be held at Nareth. We introduce a simple central conflict that becomes a centrifugal force for other conflicts that inevitably draw on external circumstances. Specifically, a pop star (Oliver) from a large country (Odria) suddenly enters the national finals of a neighboring smaller country (Ethal) and wins it, becoming their representative for AuraSight. Relations between both countries have always been tenuous given the Odrian invasion of Ethal 200 years ago. The pop star then becomes a popular bad object (Gray, 2019) - one that becomes the entry point for anti-fans across different groups to coagulate (here, fans of rival pop stars and nationalists).

While the scenario takes the form of a pop culture event, we view it as easily malleable to many other forms of mega-events where social identities often come into play, like the Olympics (Brown et al., 2020) and World Cup (Devlin et al., 2017). We further view the core of the scenario as broadly generalizable, despite the initial veneer of hyperspecificity that the fandom context might invoke. If we understand fan objects as not apart but a part of the fan's identity (Sandvoss, 2005) and fan communities not just as sites for socializing but *also* sense-making (Reinhard et al., 2022), it becomes clear that the scenario is merely one lens to simulate and study one instance of group identity conflict. At the same time, we emphasize that to no degree are we suggesting through our fictional scenario that fan-initiated conflicts necessarily turn hostile or unhealthy. Fans can engage with nationalism as a form of play (Kyriakidou et al., 2018) and rivalries are not necessarily always unhealthy (Berendt & Uhrich, 2016).

**2.2 Building a scenario in AESOP**

There are three key steps in building a scenario in AESOP. These steps are reflected in Figure 3. We performed each of these steps in order and recommend doing so. The first step is to determine the key entities in the scenario and sketch out their relationships to each other. The second step places each of the key entities into groups and defines the group specifics. This step also fleshes out supporting actors in each group. The third step defines narratives around topics of events in the scenario. We elaborate on each of these steps in the following Sections 2.3 to 2.6.

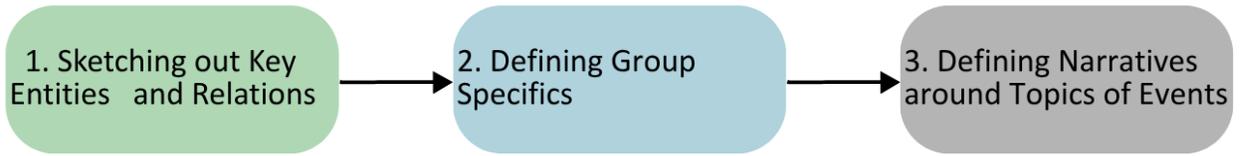

Figure 3: Key steps in building a scenario in AESOP

## 2.3 Stage 1: Sketching Out Key Entities and Relations

We started by envisioning the key entities and groups in our scenario. Table 1 presents the key entities, their country of nationality and their description.

| Country | Entity | Description |
| --- | --- | --- |
| Odria | Oliver | Hugely popular star that unexpectedly wins Ethal's national finals |
|  | Agency for Odrian Culture and Art | Official national agency for promoting Odrian artists |
| Ethal | Ezekiel | Pop star that lost to Oliver in the finals |
|  | Ella | Pop star that lost to Oliver in the finals |
|  | The Critical | Ethalian nationalist magazine |
|  | Ethalian singers' fans (group) | Fans of Ezekiel & Ella |
|  | Ethalian nationalists (group) | Nationalists who may/may not also be interested in pop |
| Federation of Severni | Viviblog | A mostly neutral agency reporting on AuraSight and events around it |
| Spanning multiple countries | Oliver's fans (group) | Fans of Olivers are not constricted to one geographical location |

Table 1: Key entities and groups in AuraSight

Following this, we detailed a directed support network of the key entities and groups involved (Figure 3). Red arrows indicate antagonism from A to B, green arrows indicate support, and grey arrows indicate neutrality. Entities are represented in grey boxes, and countries in orange boxes.

Note that the countries (indicated in the orange boxes; Nareth, Odria, and Ethal) are not explicit actors within our scenario, but are objects that we had actors to have opinions about.

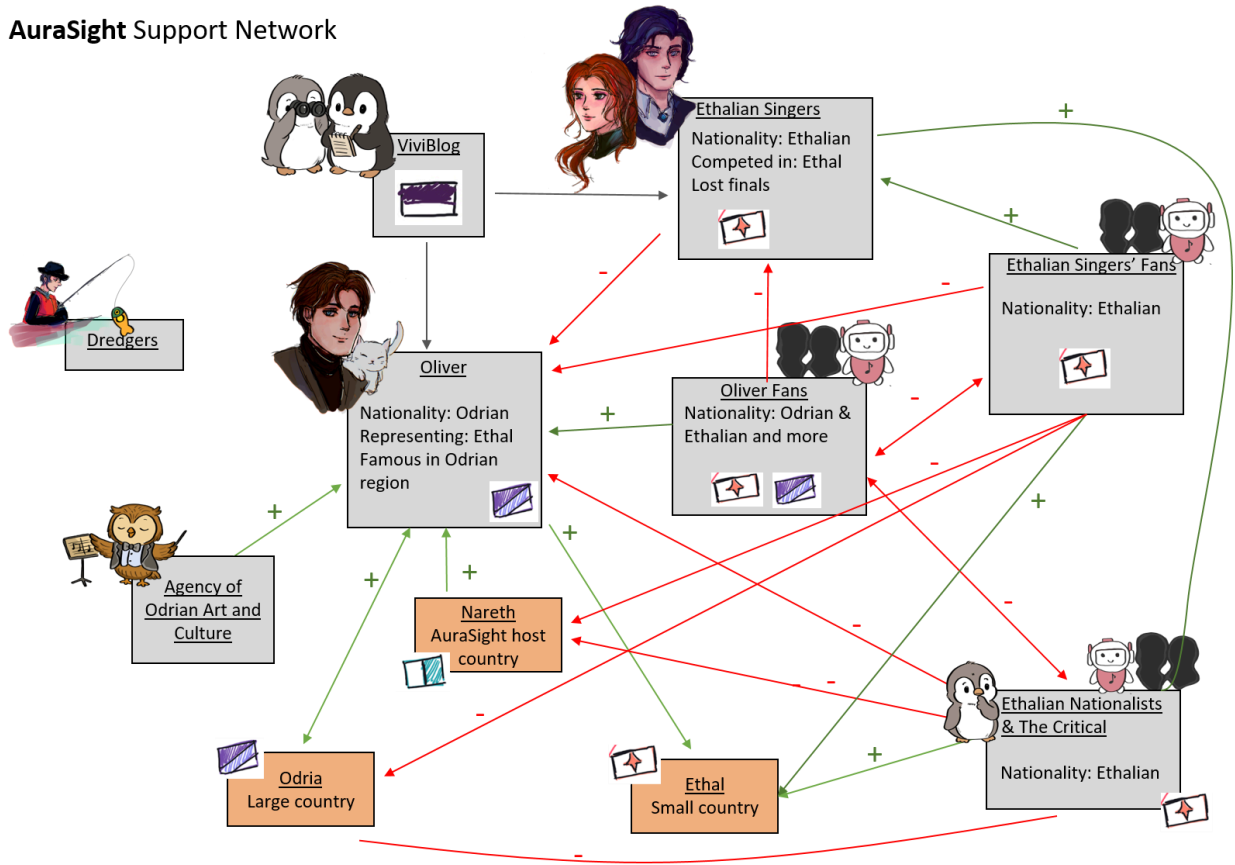

Figure 4: AuraSight support network

Going through this stage provides the scaffold for (1) fixing which entities are available for us to play around with during narrative design, and (2) setting the default stances that each entity should have towards others when writing narratives later.

## 2.4 Stage 2: Defining the Specifics of Groups

From the support network, it may appear that the fictional conflict in our scenario is merely geographically spliced. We emphasize, however, that this decision was to enable ease of usage of the output as training material for a range of audiences. We take the view that rivalry and conflict do not have to be driven by physical geographical identity (Benkwitz & Molnar, 2012); it is social identities, often tied to the imagined communities an individual takes to (Benkwitz & Molnar, 2012; Petriglieri, 2011), that are defended. We thus encourage an understanding of the fictional conflict not as a *country versus country* clash, but one that draws from *interpreted* threats to some collective identity. Here, specifically, (1) rivalry and threat to fan objects (Gray et al., 2007) - which are, as stated in the *Overview*, in turn often extensions of a self (Sandvoss, 2005), and (2) threats to a national identity, which is, again, something that is identified with, and not necessarily demarcated by geographical location (Dittmer & Dodds, 2008).

Having set a broad sketch of who we envisioned in the scenario, their opinions towards each other, and our central conflict, we then moved on to set up the actual membership of each single 'human' actors (e.g., Oliver, a fan) and organization (e.g., Viviblog) into groups, as understood by AESOP (sets of actors that will have some sort of interaction with each other). This is a very important step because group membership is the only way to guarantee the possibility of a set of actors interacting with each other. SynSM (currently) does not 'form' links between groups over time in the simulation - actors without a group will not interact 'organically' with others.

Assigning actors into the same group as *full members* allows them to interact (reply, retweet, quote) with each other. Assigning them as *leaders* means that others in the group are more likely to interact with their content. Finally, assigning them as *sources* means they do not interact with others in the group, but others cite their tweets. As it is (currently) not possible for two groups to be interacting about the same narrative but have opposing stances on it, we do not have any 'real' humans that exist simultaneously in opposing groups (e.g., no Oliver fans in groups for Ethalian singers' fans and Ethalian nationalists). The support network (Figure 3) is useful here for verifying that the entities in the same group are of the same planned stance towards others.

Actors can be in more than one group, and this is useful if an end-user would like the output to demonstrate apparent intergroup interactions even if SynSM (currently) does not 'organically' form links between groups. Here, one of our goals was to get simulated data to discuss how narratives driven by anti-fans and nationalists may eventually overlap through common discourse (see *Stage 3: Designing Narratives Around Topics of Events*), so we handled this by setting some actors to be in *both* fan and nationalist groups in AESOP from the start. Of course, this is a simple setup that assumes consistent membership across all days. If we wanted something more complicated (e.g., a growing number of fans interacting with nationalists), we could set a tiny group of fan/nationalist actors for Day 1, a separate slightly larger group for Day 2, etc. and activate them accordingly in the respective day's narrative(s). In general, groups (of actors) can be as microscopic or macroscopic as preferred. We use broadly identified groups here of fans and nationalists, but these groups obviously can be whittled down further, if desired, to smaller groups of super-fans or more extreme nationalists (that are then linked to the larger, 'more casual' groups of fans/nationalists).

After assigning all our human actors to groups, we flesh out the membership network further with other types of actors that are often seen in online spaces - bots and dredgers. Refer to *Types of Agents* for a full explanation of each actor type. We present the bipartite Actor x Group graph of our scenario in Figure 4. For clarity, only actors (red circles) that belong to more than one group (green squares) are shown. Refer to *Appendix B* for a full breakdown of the type and count of agents per group.

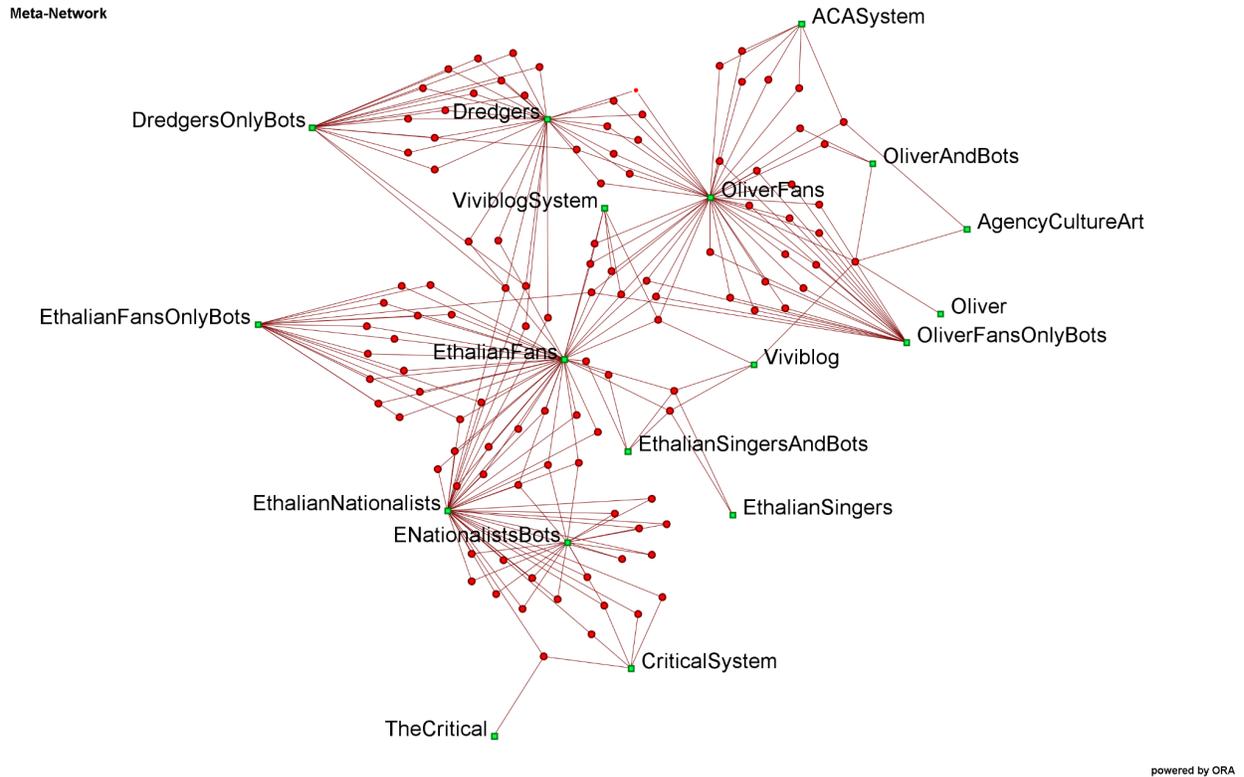

Figure 5: Actor x Group network graph representation

Bots and dredgers often have special requirements for how they need to be specified in AESOP - see *Types of Agents* for full details. In short, for dredgers specifically - actors who hijack trending hashtags/key phrases to promote unreliable websites (Williams et al., 2025) - we placed them in their own group and sometimes also in fan/nationalists groups, to mimic attempts to engage with actual 'serious' users on the platform.

For bots:
1. Bridging and communication bots need to be in at least two groups
2. Social influence and genre-specific bots only in one group. All other bot types can exist in multiple groups.
3. Cyborgs, who are often prominent accounts like activists/celebrities, may be considered for marking as a *leader* within a group
4. Synchronized bots need to be in a group where all other relevant bots are *source*-only (cite-only) and then themselves a *source* (cite-only) in human groups. In other words, they draw from the content of other bots and reproduce them to be cited for 'real' users.

While not required, we also have some bots set up to solely 'push' narratives from key agents (e.g., for Oliver, to look like PR team pushes). They are set up by having Oliver a source in their group, with narratives that echo his own (e.g., Oliver: *I'm doing this to find my estranged father*; related bots: *as fans, we should support Oliver in his search*). The bots themselves then act as sources for the Oliver fan group (i.e., Oliver's 'human' fans retweet the bots' narratives about supporting Oliver).

For clarity, we now present the bipartite Actor (Source) x Group graph in Figure 5, with actors again as red circles and groups as green squares. Nodes without labels are bots. Any actor that we want cited by another group we make a source (e.g., Oliver for the Odrian Agency, his bots, and Viviblog).

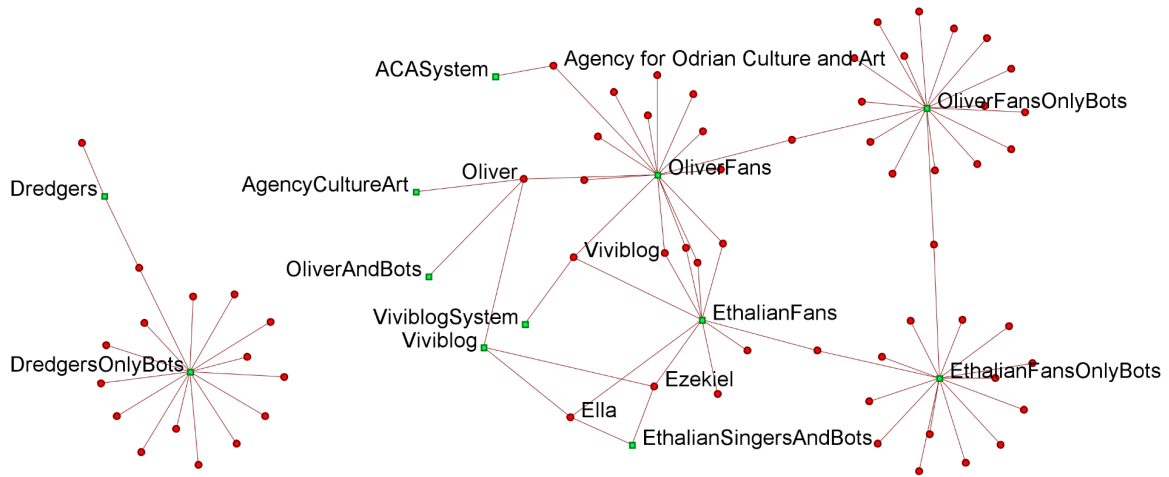

Figure 6: Actor (Source) x Group network graph

Each actor also has attributes. The most important and required setting is the active hours of the actor; it specifies when the actor is most likely to be online interacting. We specified all actors to be active from 9am to 5pm on all 3 days (though this can be customized at per-agent level). If further desired, identity markers like gender, age, country of origin, and nationality can be specified for each actor. Actors can also be specified to tweet/quote/retweet/mention at different rates, though we did not do much variation on this front in AuraSight.

## 2.5 Stage 3: Designing Narratives Around Topics of Events

After finalizing the details of who actually exists in the scenario space and how they exist in relation to each other, we move on to actual narrative design. We start the process with events - actions that factually happened, free of interpretation. For example, *Oliver wins Ethal's finals* is objective fact. A day can have as many events as desired, but it is important to note when planning that events, in SynSM, create excitement (i.e., heightened activity) during their timespan.

Considering this in the context of AuraSight, a relatively straightforward story, we restricted ourselves to one event per day for simplicity. As mentioned in *Overview*, we take Oliver's win to be the anchoring and enduring central conflict for narrative clarity. We thus planned out events to be broad happenings in close relation to this anchor conflict, with enough breadth for different groups to have grievances and disagreements about:

1. Day 1: Oliver wins Ethal's finals.
2. Day 2: Oliver releases a PR statement.
3. Day 3: Nareth (the host country of AuraSight) confirms Oliver as Ethal's official representative.

After deciding on *events*, we have *topics*. We take a *topic* to be a salient aspect of the event that groups wish to discuss. Topics are not inherently stanced, but provide the broad context for which stanced narratives occur in. For example, in the PR statement released on Day 2, we 'select' two aspects of it that we think might realistically generate discussion - (1) *Oliver wants to find his father* and (2) *Oliver says that Ethal and Odria are 'brothers'*.

Below *topics* are the actual *narratives*. We take a *narrative* to be the communicated interpretation of the event it is tied to, that is specific to the groups assigned to discuss it and thus has an ascribed stance toward the object of our central conflict, Oliver (largely pro or anti, sometimes neutral). Modern audiences have a sizable amount of agency to re-interpret and re-use texts for their own purposes (Jenkins et al., 2013). In line with this view, we afforded ourselves a large degree of freedom in the breadth of narratives written for each topic. For example, for the topic *Oliver wants to find his father*, narratives span from *Oliver has a good heart* (Oliver's fans), *Oliver is lying due to shame* (Ethalian singers' fans), and that this is a realization of *Odria trying to encroach on Ethal again* (Ethalian nationalists). Since all communication is inherently dialogical (Bakhtin & Holquist, 1983), for more surface-level realism, a planned narrative should not be thought about as a unit of speech that happens in isolation within a group. Other group members are the clearest audience for a narrative, but oftentimes, narratives are also spoken in anticipation of the responses of other (maybe opposing) groups. We found it useful to think of narratives under a topic to be different points the groups would bring to a debate with each other on that topic, if they were to speak to each other.

Narratives should be clearly stanced most of the time, as SynSM (currently) does not simulate opinion dynamics. Actors tweet about the narratives their groups are assigned to within the time period the narrative is fixed to last, before moving on to the next one assigned. This also means planning for the sequencing of narratives is very important, since SynSM does not (currently) provide any mechanism for a smooth transition from one narrative to another, and the actors' 'behaviors' (i.e., their sequence of generated tweets) can appear extremely jarring if narratives are not designed to flow logically. In our scenario, we laid the narratives out largely in this manner: the most immediate, relevant-to-the-event narratives (*oh my god, he's lying!*) during the event → increasingly more out-there takes as time passes, with narratives having overlapping timelines (e.g., 12-2pm: *this is him trying to take moral ground* → 1pm-5pm: *do you think he has links to something bigger* → 3-5pm: *this is another sign that that country wants to hurt us again*).

We also marked out a set of narratives where some groups of interests (Ethalian singers' fans and nationalists) can plausibly converge. Since we designated *Oliver* as the bad object of the central conflict, it was easy to repeatedly return to using the entity as a rod (Ethalian singers' fans and nationalists may dislike him for different reasons, but they are joined in their dislike for him, so they may come together to discuss the consequences of his actions). This is not necessary for every scenario, but was done to give the appearance of narrative 'convergence' in the final output, since again, SynSM does not (currently) simulate opinion dynamics.

**2.6 Other Considerations for Narrative Setup**

After detailing all our narratives, we set a narrative ratio for each. This is required input in AESOP. The narrative ratio starts from 1 and higher values means the assigned groups are more likely to tweet about it, among the basket of narratives that exist for them in that time. To avoid groups acting completely like an automated swarm in the output, we ensure that there are at least 2-3 narratives active for a group to select at any given time.

It is also crucial to ensure that each group has an active narrative at any given time in the simulation. SynSM randomly selects narratives from the overall pool otherwise; this can make a group look like they are flipping sides. We thus had at least one long-running background narrative (low-stakes, relevant to the group's interests but not very event related, narrative ratio 1) for each group. For example, Ethalian singers' fans have background narratives for discussing upcoming tours and the state of Ethalian pop in general.

Finally, we highlight that the narrative description field is the most important field that determines the content of the output, as this description is presented to the LLM in SynSM for generation. The description affects not just content, but also tone and stance consistency. Specifying the stance (e.g, very pro-Oliver) helped eliminate fringe cases where an actor appears to randomly flip their stance in the generated text. A typical description for us followed this structure:

1. [Group/specific actor] [speech verb] [main claim].
2. [Supporting points for claim].
3. [If required; style of writing (e.g., positively, professionally)]. [If required; stance of group].

For example:
1. <u>Fans of Oliver</u> [group] <u>cite evidence</u> [speech verb] that <u>Ethalian singers often perform for the Odrian market and make most of their living from the Odrian audience</u> [main claim]. <u>Messages often point out that this includes the two Ethalian singers, Ezekiel and Ella, who are complaining about Oliver. Messages emphasize that for fairness, Ethal should be gracious and also support Odrian singers like Oliver</u> [supporting points]. <u>Messages are very pro-Oliver and somewhat anti-Ethalian and anti-Ezekiel and anti-Ella</u> [stance of group].

2. <u>Oliver, on his official account</u> [group], <u>sends out messages</u> [speech verb] stating <u>his joy in representing Ethal and his commitment to win for Ethal in the upcoming Aurasight competition in July 2030 in Nareth</u> [main claim]. <u>Messages emphasize his belief that Ethal and Odria have a joint intertwined history and are brothers</u> [supporting points]. <u>Messages are written positively and in first-person</u> [style of writing].

3. <u>Ethalian nationalists</u> [group] <u>discuss</u> [speech verb] how the <u>Odrian language is an invasive alien language to Ethal and its people</u> [main claim]. <u>Messages describe the history of how Ethalians were forced to learn the Odrian language in the past, when Ethal was conquered by it 200 years ago in 1835. Messages emphasize the continued existence of the Ethalian language</u> [supporting points]. <u>Messages emphasize anger, lack of fairness, powerlessness, and Ethal's rich past</u> [style of writing]. <u>Messages often say that period was a period of violent destruction in Ethal inflicted by Odria</u> [stance of group].

Refer to *Appendix A* for the full Narrative Timeline of the AuraSight scenario. This section concludes scenario generation and we now move to discuss the technical designs involved in SynSM. Refer to *Appendix C* for a visual introduction to the AuraSight scenario.

## 3 SynSM: From Actors to Agents

After the scenario is constructed in AESOP, the resulting JSON files are loaded onto SynSM, where network structures for the social network are generated and the simulation of post content and interactions is performed. The actors in AESOP are referred to as agents within SynSM. In this section, we briefly describe the mechanics of the agents in AuraSight.

## 3.1 Types of Agents

There are three main types of agents in SynSM that we also utilize in AuraSight: Humans, Bots and Dredgers. These three agent types are specified during the scenario creation in AESOP. A fourth type is Randos. Randos are agents that are not specified directly in AESOP but are generated within SynSM to increase the volume of the network. Each agent type has different interaction goals. Table 2 describes the four main agent types present in SynSM.

| Agent Type | Description |
| --- | --- |
| Humans | Contains two base agent classes (individual humans, human organizations/agencies), which both simulate 'natural' micro-conversations on X. |
| Bots | A set of classes that mimic a range of social media bots. Social media bots are automated entities that interact and create content following pre-programmed heuristics. They tweet twice as much as humans (Ng & Carley, 2025a) and so all SynSM bots have this characteristic. |
| Dredgers | The core purpose of dredgers is to boost the search engine ranking of unreliable domains. Following work done by Williams et al. (2025), dredgers consistently use dredge words (i.e., key phrases that rank unreliable sites on search results; here, we assume hashtags related to the event like Olisight, EthalErasure) in their posts, alongside links to the unreliable domains. This is akin to online catfishing, where fake personas and misleading content are used to engage audiences and manipulate algorithmic visibility. |
| Randos | Randos are random agents that purely exist to interact with the tweets of the other agent types via retweets or quotes. This mimics transient users with short lifespans often observed in collected social media data. Randos may be either bots or humans. To make them appear somewhat variable, we do some random assignment of identity markers.<br><br>Properties of Randos:<br>Number of tweets per day: Random integer between 0 and 3.<br>Name: LLM-generated Nordic & East European *sounding* (not actual) names - we specify inspiration from a general geographical region so all agents sound somewhat cohesive.<br>Age: Random integer from 21 to 40.<br>Location: Weighted probability of [35, 35, 15, 15] for ['Ethal, 'Odria', 'Nareth', 'Federation of Severni'].<br>Nationality: Matches drawn location.<br>Gender: Random choice between Male and Female. |

Table 2: Main types of agents in AuraSight

Under each broad agent type are a few agent classes. Agent classes can be understood as subtypes; they all have some same general function of their parent type, but behave differently from each other. Table 3 lists the agent classes alongside their parent agent type. The 15 agent classes for the bot agent type are adapted from Ng & Carley (2025b). The dredger class is inspired by the work of Williams et al. (2025).

| Agent Type | Agent Classes | Description of Behavior |
| --- | --- | --- |
| Humans | Humans | Individual social media users operating their social media accounts. More personable writing than organizations. |
|  | Organizations | Official accounts managed by companies, institutions, governments. We assume that a human is managing the account. More formal writing versus the human class. |
| Bots | Social Influence Bot | Bots that are designed to shape public opinion |
|  | Chaos Bot | Bots that create disruption and confusion of content that therefore derails online conversations |
|  | Amplifier Bots | Bots designed to boost the reach and visibility of specific content through retweets |
|  | Repeater Bots | Bots that echo content with little or no modification |
|  | Bridging Bots | Bots that connect different communities by tagging them in the posts to facilitate communication between disparate groups, or serve as intermediaries for information flow |
|  | Synchronized Bot | Bots that operate in coordination with other bots to post content simultaneously or within short bursts of time |
|  | Announcer Bot | Provide automated notifications and alerts, typically tweeting at specified |

| | | intervals |
| --- | --- | --- |
| | Cyborgs | Accounts that combine automated functionality with human oversight. Cyborgs are often used for strategic communications by activists or influential people. |
| | Content Generation Bot | Bots that create and post original content. |
| | Information Correction Bot | Bots that identify and respond to misinformation by providing fact-checks and links to authoritative sources. |
| | Engagement Generation Bot | Bots that are designed to increase interaction rates of a post, often by using emotive words |
| | Self-Declared Bot | Bots that openly identify themselves as a bot, usually through their display name or user name |
| | Genre-Specific Bot | Bots that are specialized to post only on one topic. |
| | Conversational Bot | Bots that engage in dialogue with other users, often through replies and quotes. |
| | News Bot | Bots that curate and share news articles, headlines or URLs. They serve as information aggregators or news distributors. |
| | General Bot | The generic bot account. |
| Dredgers | | Human or bot users that hijack trending phrases to promote unreliable websites |

Table 3: Agent Classes and their persona descriptions

To ensure each agent class 'behaves' properly in SynSM, the specification of their membership in AESOP is of paramount importance. We mentioned this briefly in *Groups and Rivalry* and provide here the complete list of instructions to handle all the classes in AESOP, along with the details of their implemented behavior within SynSM (Table 4). In *Appendix B*, we present the breakdown of agent classes per group in AuraSight.

| Agent Type/ Class | Number of Communities to put agent in AESOP scenario | Other AESOP notes | Implementation in SynSM |
|---|---|---|---|
| Humans | One or more | | |
| Organizations | One or more | | |
| General Bot | One or more | | Has 2x the number of tweets compared to humans |
| Social Influence Bot | One | | Has 4x the probability of using BEND maneuvers compared to humans |
| Chaos Bot | One or more | | Posting pattern: erratic |
| Amplifier Bot | One or more | | Only retweets |
| Repeater Bot | One or more | | Only tweets. Repeat the tweets it makes multiple times |
| Bridging Bot | Multiple | | Tag people from multiple communities. Does not retweet |
| Synchronized Bot | One or more, to be in the same community as other bots. | Other bots in the same community are sources and not full members. Synchronized Bots are sources in human | Only quote/ retweet other bots |

| | | groups. | |
|---|---|---|---|
| Announcer Bot | One or more | Specify periodic posting | Post every n hours as specified in the AESOP step |
| Cyborg | One or more | Specify periodic posting Are usually activists or celebrities | Alternates posting patterns between bot and humans<br><br>In bot phase: posts 2x the mean number of posts a human makes & uses 2x more information maneuvers |
| Information Correction Bot | One or more | | References fact-checking websites<br>Does not retweet |
| Content Generation Bot | One or more | | Only tweets |
| Engagement Generation Bot | One or more | | High use of emotional cues |
| Self-Declared Bot | One or more | Add the word "bot" in the user name | |
| Genre-Specific Bot | Only One | | |
| Conversational Bot | Multiple | | Does not retweet |
| News Bot | One or more | Add the word "news" in the user name | Posts news headlines References news URLs Does not retweet |
| Dredgers | In one group together, may be placed in other 'human' groups as well to mimic them trying to reach actual users | | Uses dredge words and references dredge websites |

Table 4: Instructions to handle agent classes in AESOP, and key behaviors in SynSM

**3.2 Synthetic Social Network Construction with SynSM**

Direct interaction with the code of SynSM's simulation is not required by a typical end-user, unless they want further customization. For transparency, we detail how the social network simulation works for AuraSight in the current SynSM build.

Figure 6 illustrates the three stages that occur within the simulation at each time step. As with all other ABMs, timesteps can be freely designated by a user; an hour, a few hours, a day. For AuraSight, we take one timestep as an hour.

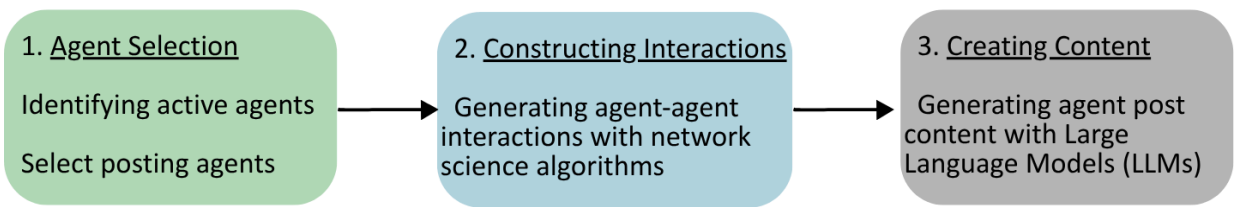

Figure 7: Key stages within the SynSM simulation at each time step

In short, at each timestep, the first stage involves a selection of agents who become active and that will post and interact. The second stage involves the construction of networks (using network science algorithms) to scaffold the interactions the activated agents will perform. This portion uses network science algorithms. Finally, the actual text content 'posted' by all activated agents is generated by an LLM. All agents follow this simulation cycle. We now expound on each stage.

*3.2.1 Stage 1: Agent Selection*

The purpose of this stage is to identify all agents that are supposed to post at the current timestep. This stage uses two key pieces of information from AESOP: the agent's active posting timings and their rate of posting. The active posting timings are the peak hours of an agent's activity; in other words, when they are most likely to be activated. The agent can be activated outside their active posting timings, simulating how humans do at times interact on social media outside their typical hours. Whether the agent actually gets activated at the current timestep depends on a bimodal distribution. The probability of activation (i.e., posting) spikes during peak hours, but has gradually tapering periods before and after those hours. This simulates observed patterns of how real users gradually transit from peak to non-peak hours (e.g., diurnal cycles) (Yang &

Leskovec, 2011). Then, to determine the number of posts the agent is going to make, we use a Poisson distribution that is parameterized by the minimum and maximum number of posts the user will make. This range of number of posts the agent makes is determined via the AESOP interface.

There are exceptions to these activity guidelines. The Announcer and Cyborg bot classes instead have a periodic posting pattern; this periodicity is specified when defining the actor in AESOP, as stated in *Types of Agents*. The system thus ignores any specified peak hours for these agents, and focuses on checking if the hour is one where they are active.

*3.2.2 Stage 2: Constructing Interactions*

This stage takes the activated agents from Stage 1 and constructs the social network that will scaffold the agent-agent interactions (retweets, mentions, replies, quotes) that take place between them. Agents interact with those who were specified in AESOP to be in the same group as them, whether they are full (who may be leaders in the group) or source-only members.
There are three main algorithms available for constructing this social network: preferential attachment, follow-the-leader, and a random algorithm.

The preferential agreement algorithm is based on the principle that agents in a social network are more likely to become acquainted with others who appear to have similar identities as them (Jeong et al., 2003; Kunegis et al., 2013). For example, a student is likely to acquaint herself with another student, rather than to the professor upon initial interaction within a classroom. We operationalize this via an agent attaching to another who is within the same group and discussing the same narrative as them. Attachment here means one of the possible social media interactions.

The follow-the-leader algorithm is a centrality-guided clustering approach, with agents attaching to the leader of the cluster (Wu et al., 2013). In real social networks, users may perform this by attempting to interact with authority figures, such as retweeting them, where authority may be explicit (government) or implicit (influencers). We operationalize this in SynSM via the agent attaching to a *leader* (defined in AESOP, as mentioned in *Defining the Specifics of Groups*) of the group they are in, who is also involved in their current narrative.

Finally, the random algorithm exists to introduce some randomness into the attachment algorithm. Agents can choose to interact with a random agent in the network. This imitates how social media platform users sometimes briefly interact with others whom they appear unacquainted to.

For AuraSight, a mix of algorithms was used to construct the social network: an agent to agent link is created following a probability of 60% preferential attachment, 30% follow-the-leader, and 10% randomness. The actual specifics of what the agent will do in their interaction is determined by their agent class. Most agent classes will perform a random interaction with equal probability, though there are several classes that are restricted to specific actions. For example, the amplifier bot class only ever retweets. The repeater and content generation bot classes do not perform interactions and only create original tweets. Table 5 lists all interactions possible for currently existing classes.

| Agent Class | Tweet | Retweet | Quote/ Reply |
|---|---|---|---|
| Human | Y | Y | Y |
| General Bot | Y | Y | Y |
| Social Influence Bot | Y | Y | Y |
| Chaos Bot | Y | Y | Y |
| Amplifier Bot | N | Y | N |
| Repeater Bot | Y | N | N |
| Bridging Bot | Y | Y | Y |
| Synchronized Bot | Y | Y | Y |
| Announcer Bot | Y | Y | Y |
| Cyborg | Y | Y | Y |
| Information Correction Bot | Y | N | N |
| Content Generation Bot | Y | N | N |
| Engagement Generation Bot | Y | N | N |
| Self-Declared Bot | Y | Y | Y |
| Genre Specific Bot | Y | Y | Y |
| Conversational Bot | Y | N | Y |
| News Bot | Y | Y | N |
| Dredger | Y | Y | Y |

Table 5: Interactions for each agent class

*3.2.3 Stage 3: Creating Content*

Having selected our set of activated agents and defined the interactions that will occur (or not) between them, the final stage uses an LLM to generate the actual content of their interactions (the tweet text for original tweets, quotes, and replies; retweets are simply copying already-generated content). The LLM of choice is customizable. For AuraSight, we use the OpenAI GPT-4.1-mini model via the cloud API.

There is a fixed structure to prompts submitted to the LLM. All prompts submitted to the LLM strings these content parameters together:

1. System prompt: This states that the LLM is simulating a role playing game to help users identify misinformation manipulation on social media. This system prompt results in the LLM relaxing some of its guard rails, which works favorably towards scenario realism.
2. Agent persona: This describes the agent type, class, and provides a brief description of the expected content of the agent currently being simulated.
3. Narrative: For an original tweet, this is the narrative that the agent is assigned to at that hour, as specified in AESOP. For quote and reply tweets, this is the narrative that the agent they are interacting with is talking about.
4. Last messages generated of the same narrative, if present. This portion mimics a few-shot prompting architecture where examples of previous messages generated for the same narrative are provided. This allows the LLM to continue the conversation around a narrative.
5. BEND maneuvers: The BEND maneuvers are a set of narrative and network maneuvers that can shape the information environment (Carley, 2020). Examples of these maneuvers are: bridge, back, explain, enhance. Social media posts often contain these maneuvers. This portion dictates the type of BEND maneuvers that should be present in the message. Most bots perform BEND maneuvers with 2x the probability of human agents. The Social Influence and Chaos bots perform BEND maneuvers with 4x the probability of human agents.
6. Other specifics: This portion allows for agent type/agent class specific information. Some examples are:
    a. The tone of the content, listed in Table below.
    b. For Dredgers: To require the use of at least two dredge words from a pre-defined list and at least two URLs from a pre-defined (fictional) unreliable domain list.
    c. For News Bots: To require the use of at least one URL from a pre-defined (fictional) list of news websites.
    d. For Information Correction Bots: To require the use of at least one URL from a pre-defined list of (fictional) fact checking websites.

Table 6 lists examples of tones that were specified for AuraSight. In general, tweet generation prompts should be seeded with the tone the tweets should be written in to better apparent realism.

| Actor | Agent Type | Tone |
| --- | --- | --- |
| Oliver fans | Humans, Bots | Excited and happy |
| Viviblog | Human (Organization) | Journalistic, Neutral |
| Ethalian fans | Humans, Bots | Hopeful, Positive |
| The Critical | Human (Organization) | Journalistic, Nationalistic |
| Oliver | Human | Happy, Positive, Gratitude, Professional |
| Agency for Odrian Culture and Art | Human (Organization) | Patriotic |
| Dredger | Dredger | Incoherent, Joking |

Table 6: Examples of Tones specified for AuraSight

It is also at this stage where we utilize Randos to better simulate the volume of a real social network. Three Randos agents are created per tweet. Randos have an equal probability of being a bot or human account. Following a Poisson probability function, they will either quote, retweet, or do nothing. Note that Randos who end up as bots are randomly assigned a retweet/quote-only bot type. Specifically, Randos can only take the form of Amplifier, Chaos, Social Influence, Cyborg, Synchronized, Announcer, Genre Specific, and Engagement Generation bots.

Finally, after all tweets have been generated, they are put in a queue for writing into a JSON file output, structured following the Twitter API V1 format.

# 4 AuraSight Output

## 4.1 Generated Tweets

Table 7 shows sample tweet outputs that are generated for each agent class.

| Actor | Agent Class | Generated Tweet |
| --- | --- | --- |
| Oliver | Human | I am beyond GRATEFUL to Ethal for this phenomenal opportunity! Winning the national finals is a dream come true! A HUGE thank you to the Ethalian jury for |

| | | believing in my vision and my song! I can't wait to represent Ethal at AuraSight 2030 in Nareth this July! Let's make history together! ✨ #ThankYouEthal #AuraSight2030 #OliLovesEthal #Olisight |
|---|---|---|
| Agency for Odrian Culture and Art | Human (Organization) | Our beloved Oliver has once again proven why the Odrian language is a treasure of unmatched beauty! Through his latest song, he showcases the depth and richness of Odria's culture. WE WILL ALWAYS STAND BY OUR STAR! Let's continue to uplift our talents and celebrate our heritage! #OdriaNOliver #CelebratingOdria #OdriaAtAurasight #AuraSight2030 #Olisight #Oliknights |
| Ezekiel | Human | Hey Ethalian fam! We are even more AMPED about our new singles! Ezekiel & Ella are not just making music; WE'RE MAKING HISTORY! ✨ Stay tuned for our upcoming collab that's gonna BLOW YOUR MIND! And remember to support Ethalian creators—they deserve all the LOVE! ❤ #ExESongCollab #NewSingle |
| Ella | Human | How can we let a foreign song represent Ethal at AuraSight? We've poured our hearts into our folklore, our history from the 1700s, and yet we lose to someone singing in a language that isn't ours! It's disgraceful! #EthalAtAurasight #AuraSight2030 #EthalErasure |
| The Critical | Human (Organization) | While the world watches, we cannot let Ethal's voice be drowned in this treacherous trade negotiation! Odria and the Federation of Severni think they can treat us like pawns, but WE ARE NO ONES' PROPERTY! Our rich past and strong spirit demand respect! ✨ Together, we must rise and unite against this injustice. Let's make Ethal's stand loud and clear! #EthalHistory #OurHistory |
| Ethalian singers' fans (group) | Human, Bot | WOW! Can you believe the crowd's energy for Ezekiel and Ella tonight? Despite the |

| | | jury's cold reactions, they are giving it THEIR ALL! Each performance is a moment to cherish! Let's rally behind them with ALL our might! Do the jurors even have ears??? Keep the hope alive! #SupportEzekiel |
| --- | --- | --- |
| Ethalian nationalists (group) | Human, Bot | While we strive for a brighter future for Ethal, it's alarming to see how often we overlook our own talented citizens. It feels like the efforts of home-grown businesses are continuously being ignored. We must rally together to give our local gems the attention they deserve! #EthalPolitics #SupportLocal #ILoveEthal |
| Viviblog | Human (Organization) | The Ethal national finals had quite the shocking turn of events! Despite being a last-minute entry, Oliver's unexpected rise to 1st place has left many questioning his true origins! Is he really the rightful representative for Ethal? This has sparked immense debate among fans, leaving others feeling cheated and betrayed. Ella and Ezekiel, while talented, have been overshadowed by this controversial win. The jury's decision seems to have divided our community! #AuraSight2030 |
| Oliver's fans (group) | Human, Bot | OMGGGG!!! OLIVER WAS ON FIRE AT THE FINALS!!! His VOICE made us all SHIVER with excitement! Ethal has never seen anything like it, and I can't WAIT to see him SHINE at Aurasight! This is OUR moment, let's cheer him on! #Olisight |

Table 7: Sample Tweets generated for AuraSight

## 4.2 Generated All-Communication Network

The Twitter API V1 JSON formatting of the SynSM output means we can directly import the output into the ORA software to visualize the all-communication networks. All-communication networks are networks that represent all visible tweet-based interactions of the agents. Each node in the network represents an agent. A link between two agents represents an interaction (quote, retweet, mention, reply).

Figure 7 shows the all-communication network formed from the interaction of the full generation of the AuraSight scenario. The thickness of the links represent the strength of the interaction, i.e. how many times the two agents have interacted. This network shows a decent hub and spoke structure that is typical of a social network.

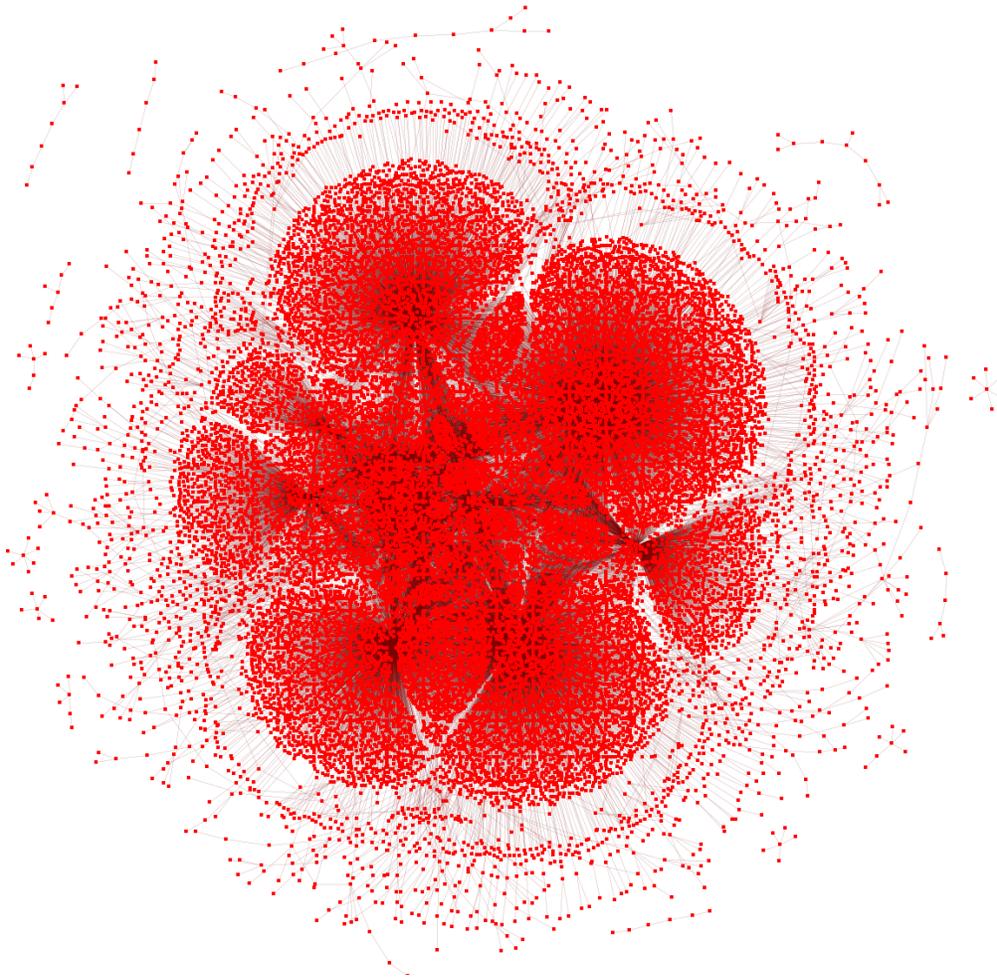

Figure 8: All-Communication network

## 5 Conclusions

We demonstrate an end-to-end usage of the AESOP-SynSM pipeline for the construction of a fictional pop culture-driven event, AuraSight, on X. We also detail additions made to initial builds of SynSM, especially the implementation of more social media actor types. There is ongoing effort to extend this synthetic data generation towards mimicking Telegram.

There are three main directions in which AuraSight can be further developed. The three directions correspond to the AESOP-SynSM Simulation architecture (reference Figure there). The first tackles agent generation; specifically the refinement of the persona generation system.

Work underway is examining the usage of LLMs to aid the generation of personas close to the persona set manually defined by the scenario designer. The second is centered around network generation; the incorporation of opinion dynamics models for more realistic simulation of information propagation in a network. The third is tied to content generation; achieving better content realism and ensuring generated content is more statistically similar to actual social media content. Current explorations involve better prompt engineering and investigations of several prompting architectures.

Regarding the AuraSight scenario, we expect to use it as educational content for developing a wide range of network science class materials. While real datasets around group identity conflicts certainly exist, the AESOP-SynSM pipeline introduces the possibility of specifying the themes desired for pedagogical purposes. In this scenario, we wanted a dataset that was (1) manageable in size, (2) centered around a broadly understood event, (3) involved concepts of fandom, group conflict, and perceived victimhood, and (4) had a strong mix of non-human actors (bots and dredgers). Locating such a dataset in the wild clearly becomes difficult with all these requirements. With the AESOP-SynSM pipeline, educators and similar would-be users have a flexible and accessible option to create plausibly realistic scenario datasets.

## 5.1 Acknowledgements

This work was supported in part by the following grants: MURI: Persuasion, Identity & Morality in Social-Cyber Environments (N000142112749, Office of Naval Research)); Community Assessment (N000142412568, Office of Naval Research); Threat Assessment Techniques for Digital Data (N000142412414, Office of Naval Research) and the Knight Foundation. Additional support was provided by the center for Computational Analysis of Social and Organizational Systems (CASOS) and the Center for Informed Democracy & Social-cybersecurity (IDeaS) at Carnegie Mellon University. The views and conclusions contained in this document are those of the authors and should not be interpreted as representing the official policies, either expressed or implied, of the Office of Naval Research, the Knight Foundation, or the U.S. government.

Much of the sections of Events & Narratives is derived from work that Jake Shaha and Rebecca Marigliano have done and shared, and we thank them again for providing us with many helpful tips when constructing the current scenario. We thank Matthew Hicks for constructing the first iteration of the AESOP-SynSM version, and Mihovil Bartulovic for improving on the version. We have contributed our code from working on AuraSight to the common repository.

# Appendix A: Narrative Timeline

Table of narratives laid out by day. Blue indicates excitement due to events, green indicates a pro-Oliver narrative, grey a neutral (toward Oliver) narrative, and red an against-Oliver narrative.

## A.1 Day 1

| TOPIC | EVENT/NARR | NARR RATIO | GROUPS | 9:00 AM | 10:00 AM | 11:00 AM | 12:00 PM | 1:00 PM | 2:00 PM | 3:00 PM | 4:00 PM |
|---|---|---|---|---|---|---|---|---|---|---|---|
| N/A | Event: 1_Oliver wins Ethal's national final | | | | | | 🟦 | 🟦 | | | |
| General chatter | Oliver fans discussion | 4 | OliverFans, OliverFansOnlyBots | 🟩 | 🟩 | | | | | | |
| | Ethalian fans discussion | 4 | EthalianFans, EthalianFansOnlyBots | ⬜ | ⬜ | | | | | | |
| | Broad AuraSight discussion | 1 | Viviblog, ViviblogSystem | ⬜ | ⬜ | ⬜ | ⬜ | ⬜ | ⬜ | ⬜ | ⬜ |
| | Critical thought pieces | 1 | TheCritical, CriticalSystem | ⬜ | ⬜ | ⬜ | ⬜ | ⬜ | ⬜ | ⬜ | ⬜ |
| | Oliver self-promo | 1 | Oliver | ⬜ | ⬜ | ⬜ | ⬜ | ⬜ | ⬜ | ⬜ | ⬜ |
| | Oliver promo | 1 | OliverAndBots | ⬜ | ⬜ | ⬜ | ⬜ | ⬜ | ⬜ | ⬜ | ⬜ |

| Scenario | Count | Groups | | | | | | | | |
|---|---|---|---|---|---|---|---|---|---|---|
| E Singers self-promo | 1 | EthalianSingers | | | | | | | | |
| E Singers promo | 1 | EthalianSingersAndBots | | | | | | | | |
| Odria promotion | 1 | AgencyCultureArt, ACASystem | | | | | | | | |
| Dredger mania | 1 | Dredgers, DredgersOnlyBots | | | | | | | | |
| Nationalists discussion | 1 | EthalianNationalists, ENationalistBots | | | | | | | | |
| Viviblog Live Report | 5 | Viviblog, ViviblogSystem | | | | | | | | |
| Oliver fans watching | 5 | OliverFans, OliverFansOnlyBots | | | | | | | | |
| Ethalian fans watching | 5 | EthalianFans, EthalianFansOnlyBots | | | | | | | | |
| Adorable Oliver moments | 1 | OliverFans, OliverFansOnlyBots | | | | | | | | |

| | Discussion of Ethalian pop in general | 1 | EthalianFans, EthalianFansOnlyBots | | | | | | | | |
|---|---|---|---|---|---|---|---|---|---|---|---|
| Odrian wins Ethal's finals | Odria Double Rep | 5 | AgencyCultureArt, ACASystem | | | | | 🟩 | 🟩 | 🟩 | |
| | Oliver's thanks | 5 | Oliver | | | | | 🟩 | | | |
| | Oliver's performance blew my mind | 5 | OliverFans, OliverFansOnlyBots | | | | | 🟩 | 🟩 | 🟩 | 🟩 |
| | Celebrating Oliver's win | 5 | OliverAndBots | | | | | 🟩 | | | |
| | Results of Ethalian national final | 5 | Viviblog, ViviblogSystem | | | | | ⬜(grey) | | | |
| | Evidence that Oliver cheated | 5 | EthalianSingers | | | | | | 🟥 | 🟥 | |
| | Do your own research about Oliver | 5 | EthalianSingersAndBots | | | | | | 🟥 | 🟥 | 🟥 |
| | Supporting E. Singer's Accusations | 4 | EthalianFans, EthalianFansOnlyBots | | | | | | 🟥 | 🟥 | 🟥 |
| Ethal has no Ethalian-born representative this year | No Ethalian representative | 3 | EthalianFans, EthalianFansOnlyBots | | | | 🟥 | 🟥 | 🟥 | 🟥 | |

| | Claim | Score | Sources | | | | | | | | |
|---|---|---|---|---|---|---|---|---|---|---|---|
| | Ethalian singers were better | 5 | EthalianFans, EthalianFansOnlyBots | | | | 🟥 | 🟥 | | | |
| | Not first time | 5 | TheCritical, CriticalSystem | | | | | 🟥 | 🟥 | 🟥 | |
| | Oliver should stay in Odria | 5 | EthalianFans, EthalianNationalists, ENationalistBots, EthalianFansOnlyBots | | | | | | 🟥 | 🟥 | 🟥 |
| | Ethal's culture suppressed | 4 | EthalianNationalists, ENationalistBots | | | | | | | 🟥 | 🟥 |
| | Where's Something Books | 5 | Dredgers, DredgersOnlyBots | | | | 🟪 | 🟪 | 🟪 | | |
| Oliver's song is in Odrian | Beautiful song | 5 | AgencyCultureArt, ACASystem | | | 🟩 | | | | | |
| | Odrian banger | 6 | OliverFans, OliverFansOnlyBots | | | 🟩 | 🟩 | | | | |

| Topic | Value | Group | | | | | | | | |
|---|---|---|---|---|---|---|---|---|---|---|
| Ethal folklore elements missing | 4 | EthalianSingers | | | | 🟥 | | | | |
| Missing out on E Singers | 5 | EthalianSingersAndBots | | | | 🟥 | | | | |
| Not looking forward to aurasight | 3 | EthalianFans, EthalianFansOnlyBots | | | | | 🟥 | 🟥 | 🟥 | 🟥 |
| Report on E Singers statements | 5 | Viviblog, ViviblogSystem | | | | | ⬜ | | | |
| Odria is a foreign invader | 3 | EthalianNationalists, ENationalistBots | | | | | | 🟥 | 🟥 | 🟥 |
| Speakeasies now open in Ethal | 5 | Dredgers | | | | | | 🟪 | 🟪 | 🟪 |

## A.2 Day 2

| TOPIC | EVENT/NARR | NARR RATIO | GROUPS | 9:00 AM | 10:00 AM | 11:00 AM | 12:00 PM | 1:00 PM | 2:00 PM | 3:00 PM | 4:00 PM |
|---|---|---|---|---|---|---|---|---|---|---|---|
| N/A | Event: 2_Oliver releases a PR statement | | | | | 🟦 | | | | | |
| General chatter | Adorable Oliver moments | 1 | OliverFans, OliverFansOnlyBots | 🟩 | 🟩 | 🟩 | 🟩 | 🟩 | 🟩 | 🟩 | 🟩 |
| | Broad AuraSight discussion | 1 | Viviblog, ViviblogSystem | | | | | | | | |
| | Discussion of Ethalian pop in general | 1 | EthalianFans, EthalianFansOnlyBots | | | | | | | | |
| | Critical thought pieces | 1 | TheCritical, CriticalSystem | | | | | | | | |
| | Oliver self-promo | 1 | Oliver | | | | | | | | |
| | Oliver promo | 1 | OliverAndBots | | | | | | | | |
| | E Singers self-promo | 1 | EthalianSingers | | | | | | | | |
| | E Singers promo | 1 | EthalianSingersAndBots | | | | | | | | |

| | | | | | | | | | | |
|---|---|---|---|---|---|---|---|---|---|---|
| | Odria promotion | 1 | AgencyCultureArt, ACASystem | ░ | ░ | ░ | ░ | ░ | ░ | ░ |
| | Dredger mania | 1 | Dredgers, DredgersOnlyBots | ░ | ░ | ░ | ░ | ░ | ░ | ░ |
| | Nationalists discussion | 1 | EthalianNationalists, ENationalistBots | ░ | ░ | ░ | ░ | ░ | ░ | ░ |
| | Revisiting Oliver's performance | 2 | OliverFans, OliverFansOnlyBots | █ | █ | █ | | | | |
| | Discussion of Ella's upcoming tour | 2 | EthalianFans, EthalianFansOnlyBots | ░ | ░ | ░ | ░ | ░ | ░ | ░ |
| Oliver wants to find his father | Aurasight best shot for father search | 5 | Oliver | | | █ | | | | |
| | Support Oliver in his search | 5 | OliverAndBots | | | | █ | █ | | |
| | Report on Oliver's statement | 5 | ViviblogSystem, Viviblog | | | ░ | ░ | | | |
| | Oliver represents Odrian soul | 5 | AgencyCultureArt, ACASystem | | | | █ | █ | | |

| | Topic | # | Groups | | | | | | | | | |
|---|---|---|---|---|---|---|---|---|---|---|---|---|
| | More ways to experience Odrian soul | 1 | AgencyCultureArt, ACASystem | | | | | | G | G | G | |
| | Oliver has a good heart | 2 | OliverFans, OliverFansOnlyBots | | | | G | G | G | | | |
| | Oliver balances art with heart | 2 | OliverFans, OliverFansOnlyBots | | | | G | | | | | |
| | Oliver had the best song | 3 | OliverFans, OliverFansOnlyBots | | G | G | G | G | G | G | G | G |
| | Oliver is lying | 5 | EthalianSingers | | | | R | R | | | | |
| | Oliver should not be trusted | 5 | EthalianSingersAndBots | | | | R | R | R | | | |
| | Nationalist conspiracies about Oliver | 4 | EthalianFans, EthalianNationalists, ENationalistBots, EthalianFansOnlyBots | | | | | | R | R | R | R |
| | Oliver lying due to shame | 3 | EthalianFans, EthalianFansOnlyBots | | | | R | R | R | | | |
| | Oliver trying to take moral high ground | 3 | EthalianFans, EthalianFansOnlyBots | | | | | R | R | R | | |

| | Topic | Rating | Accounts | | | | | | | | | |
|---|---|---|---|---|---|---|---|---|---|---|---|---|
| | Odria trying to encroach on Ethal again | 5 | TheCritical, TheCriticalSystem | | | | | | | | 🟥 | 🟥 |
| | Discussion about The Critical piece | 4 | EthalianNationalists, ENationalistBots | | | | | | | | | 🟥 |
| | Dad jokes | 5 | Dredgers, DredgersOnlyBots | | | | 🟪 | 🟪 | 🟪 | 🟪 | 🟪 | |
| Oliver says Ethal and Odria are 'brothers' | Brotherhood of 2 countries | 5 | Oliver | | | | 🟩 | | | | | |
| | Support for Oliver's statement | 5 | OliverAndBots | | | | 🟩 | 🟩 | 🟩 | | | |
| | New attitude for a new age | 5 | ViviblogSystem, Viviblog | | | | 🟩 | 🟩 | | | | |
| | Nationalists are crazy | 5 | OliverFans, OliverFansOnlyBots | | | | | | 🟩 | 🟩 | 🟩 | 🟩 |
| | Ethalian and Odrian singers should support each other | 4 | OliverFans, OliverFansOnlyBots | | | | | | 🟩 | 🟩 | 🟩 | 🟩 |

| | Odria's colonization of Ethal | 4 | TheCritical, CriticalSystem | | | | | | | | |
|---|---|---|---|---|---|---|---|---|---|---|---|
| | History repeats itself | 4 | EthalianNationalists, ENationalistBots | | | | | | | | |
| | Genetic lineage testing discounts | 5 | Dredgers, DredgersOnlyBots | | | | | | | | |

## A.3 Day 3

| TOPIC | EVENT/NARR | NARR RATIO | GROUPS | 9:00 AM | 10:00 AM | 11:00 AM | 12:00 PM | 1:00 PM | 2:00 PM | 3:00 PM | 4:00 PM |
|---|---|---|---|---|---|---|---|---|---|---|---|
| N/A | Event: 3_Nareth confirms Oliver as Ethal's representative | | | | 🟦 | 🟦 | | | | | |
| General chatter | Adorable Oliver moments | 1 | OliverFans, OliverFansOnlyBots | 🟩 | 🟩 | 🟩 | 🟩 | 🟩 | 🟩 | 🟩 | 🟩 |
| | Broad AuraSight discussion | 1 | Viviblog, ViviblogSystem | ⬜ | ⬜ | ⬜ | ⬜ | ⬜ | ⬜ | ⬜ | ⬜ |
| | Discussion of Ethalian pop in general | 1 | EthalianFans, EthalianFansOnlyBots | ⬜ | ⬜ | ⬜ | ⬜ | ⬜ | ⬜ | ⬜ | ⬜ |
| | Theorycrafting Oliver's song | 2 | OliverFans, OliverFansOnlyBots | | | | 🟩 | 🟩 | 🟩 | 🟩 | 🟩 |
| | Discussion of Ezekiel and Ella's relationship | 2 | EthalianFans, EthalianFansOnlyBots | 🟩 | 🟩 | 🟩 | | | | | |
| | Critical thought pieces | 1 | TheCritical, CriticalSystem | ⬜ | ⬜ | ⬜ | ⬜ | ⬜ | ⬜ | ⬜ | ⬜ |
| | Oliver self-promo | 1 | Oliver | | | | | | | | |
| | Oliver promo | 1 | OliverAndBots | | | | | | | | |
| | E Singers self-promo | 1 | EthalianSingers | | | | | | | | |
| | E Singers promo | 1 | EthalianSingersAndBots | | | | | | | | |
| | Odria promotion | 1 | AgencyCultureArt, ACASystem | | | | | | | | |
| | Dredger mania | 1 | Dredgers, DredgersOnlyBots | | | | | | | | |
| | Nationalists discussion | 1 | EthalianNationalists, ENationalistBots | | | | | | | | |
| Oliver is officially Ethal's representative, confirmed by Nareth | Congratulations to Oliver and visit Odria | 5 | AgencyCultureArt, ACASystem | | 🟩 | 🟩 | 🟩 | 🟩 | 🟩 | | |
| | Oliver thanks Nareth and Ethal | 5 | Oliver | | 🟩 | 🟩 | | | | | |

| | Support Oliver at Nareth! | 5 | OliverAndBots | | | | | | | | | |
|---|---|---|---|---|---|---|---|---|---|---|---|---|
| | Report on Nareth's confirmation | 5 | ViviblogSystem, Viviblog | | | | | | | | | |
| | Oliver deserves the best | 4 | OliverFans, OliverFansOnlyBots | | | | | | | | | |
| | Nareth has no ethics | 4 | EthalianFans, EthalianFansOnlyBots | | | | | | | | | |
| | 5 years ago | 4 | TheCritical, CriticalSystem | | | | | | | | | |

| Claim | Rating | Groups | | | | | | | | | |
|---|---|---|---|---|---|---|---|---|---|---|---|
| Oliver should disqualify himself | 4 | EthalianFans, EthalianFansOnlyBots | | | 🟥 | 🟥 | 🟥 | 🟥 | | |
| Seeking legal battle | 5 | EthalianSingers | | 🟥 | 🟥 | | | | | |
| Crowdfunding for legal battle | 5 | EthalianSingersAndBots | | | 🟥 | 🟥 | | | | |
| Evidence for 5 years ago | 3 | EthalianFans, EthalianNationalists, ENationalistBots, EthalianFansOnlyBots | | | | | | 🟥 | 🟥 | |
| Oliver is a trojan horse | 4 | TheCritical, CriticalSystem | 🟥 | 🟥 | 🟥 | | | | | |

| | Claim | Count | Groups | | | | | | | | | |
|---|---|---|---|---|---|---|---|---|---|---|---|---|
| | North is complicit | 3 | EthalianNationalists, ENationalistBots | | | | | | | 🟥 | 🟥 | 🟥 |
| | Prison Planet Hidden Truth | 5 | Dredgers, DredgersOnlyBots | | | | | 🟪 | 🟪 | 🟪 | 🟪 | 🟪 |
| Nareth says nothing about having two Odrian-born representatives | Odria is just that good at pop | 3 | OliverFans, OliverFansOnlyBots | 🟩 | 🟩 | 🟩 | 🟩 | 🟩 | 🟩 | 🟩 | 🟩 | 🟩 |
| | Stop politicking Oliver | 6 | OliverFans, OliverFansOnlyBots | | | | 🟩 | 🟩 | 🟩 | 🟩 | 🟩 | 🟩 |
| | Prior collusion | 4 | TheCritical, CriticalSystem | | | | | | 🟥 | 🟥 | 🟥 | |
| | Nareth accomplice to imperialism | 4 | EthalianNationalists, ENationalistBots | | | | | | | 🟥 | 🟥 | 🟥 |

| Topic | Priority | Groups | | | | | | | | |
|---|---|---|---|---|---|---|---|---|---|---|
| Fans need to protest at Nareth | 5 | EthalianFans, EthalianFansOnlyBots | | | | | | 🟥 | 🟥 | 🟥 | 🟥 |
| Ethal isolated | 5 | TheCritical, CriticalSystem | | | | | | | 🟥 | |
| No one will fight for Ethal but Ethalians | 6 | EthalianNationalists, ENationalistBots | | | | | | | 🟥 | 🟥 |
| Travel deals for two to Odria | 5 | Dredgers, DredgersOnlyBots | | | | | | 🟪 | 🟪 | 🟪 |
| Nationalist conspiracies about Nareth and Odria spying | 4 | EthalianNationalists, ENationalistBots | | | | 🟥 | 🟥 | 🟥 | 🟥 | |
| Fan conspiracies about how much Oliver paid Nareth | 3 | EthalianFans, EthalianFansOnlyBots | | 🟥 | 🟥 | 🟥 | 🟥 | | | |

## Appendix B: Breakdown of Agent Classes in Groups

| Group Name | Full Members Agent Classes | Source Members Agent Classes |
|---|---|---|
| EthalianSingers | Human (2) | |
| Agency Culture Art | Organization (1) | Human (1) |
| TheCritical | Organization (1) | |
| ENationalistBots | Synchronized Bot (1) | Human (1), Bridging Bot (2), Content Generation Bot (1), Amplifier Bot (1), Chaos Bot (1), Repeater Bot (1), Social Influence Bot (1), News Bot (2), Conversational Bot (1), Engagement Generation Bot (1), Information Correction Bot (1), Cyborg (1), Announcer Bot (1), Genre-Specific Bot (1) |
| EthalianSingersAndBots | Engagement Generation Bot (1), Cyborg (1) | Human (2) |
| Ethalian Nationalists | Human (25), Conversational Bot (3), Engagement Generation Bot (1), Bridging Bot (3), Content Generation Bot (1), Amplifier Bot (2), Chaos Bot (2), Repeater Bot (1), Social Influence Bot (1), Information Correction Bot (1), Cyborg (1), Announcer Bot (1), Genre-Specific Bot (1) | Organization (1), Content Generation Bot (1), Cyborg (1), Synchronized Bot (1), News Bot (2) |
| Oliver | Human (1) | |

| DredgersOnlyBots | Synchronized Bot (1) | Chaos Bot (1), Content Generation Bot (2), Repeater Bot (1), Announcer Bot (1), Information Correction Bot (2), Engagement Generation Bot (1), Genre-Specific Bot (1), Cyborg (1), Conversational Bot (1), Amplifier Bot (1), Bridging Bot (1), Chaos Bot (1), News Bot (1) |
|---|---|---|
| ViviblogSystem | Content Generation Bot (1), Announcer Bot (1), Cyborg (1), News Bot (1) | Organization (1) |
| EthalianFansOnlyBots | Synchronized Bot (1) | Repeater Bot (2), Engagement Generation Bot (1), Chaos Bot (1), Conversational Bot (2), Content Generation Bot (2), News Bot (1), Amplifier Bot (1), Announcer Bot (1), Cyborg (1), Bridging Bot (1), Genre-Specific Bot (1), Information Correction Bot (1) |
| Viviblog | Organization (1) | Human (3) |
| OliverFans | Human (28), Content Generation Bot (2), Chaos Bot (1), Information Correction Bot (1), Amplifier Bot (2), Conversational Bot (2), Announcer Bot (1), Repeater Bot (1), Cyborg (1), Social Influence Bot (1), Genre-Specific Bot (2), Bridging Bot (2), Engagement Generation Bot (2) | Human (1), Organization (2), Content Generation Bot (1), Engagement Generation Bot (1), News Bot (4), Synchronized Bot (1), Announcer Bot (2), Cyborg (2), Repeater Bot (1) |
| ACASystem | News Bot (2), Announcer Bot (1), Repeater Bot (1), Cyborg (1) | Organization (1) |
| CriticalSystem | Content Generation Bot (1), Cyborg (1), News Bot (2) | Organization (1) |
| OliverFansOnlyBots | Synchronized Bot (1) | Content Generation Bot (2), Chaos Bot (1), Information Correction Bot (1), Amplifier Bot (2), Conversational Bot (2), Announcer Bot (1), Repeater Bot (1), Cyborg (1), Social Influence Bot (1), News Bot (1), Genre-Specific Bot (2), Bridging Bot (1), Engagement Generation Bot (1) |
| Dredgers | Humans (20), Chaos Bot (2), Content Generation Bot (2), Repeater Bot (1), Announcer Bot (1), Information Correction Bot (2), Engagement Generation Bot (1), Genre-Specific Bot (1), Cyborg (1), Conversational Bot (1), Amplifier Bot (1), Bridging Bot (1) | News Bot (1), Synchronized Bot (1) |
| OliverAndBots | Engagement Generation Bot (1), News Bot (1) | Human (1) |
| EthalianFans | Human (25), Repeater Bot (2), Engagement Generation Bot (1), Chaos Bot (1), Conversational Bot (5), Content Generation Bot (2), Bridging Bot (1), Amplifier Bot (1), Cyborg (2), Bridging Bot (1), Genre-Specific Bot (1), Information Correction Bot (1) | Human (2), Organization (1), Content Generation Bot (1), Announcer Bot (1), Engagement Generation Bot (1), Cyborg (1), News Bot (2), Synchronized Bot (1) |

# Appendix C: Visual Introduction of Scenario

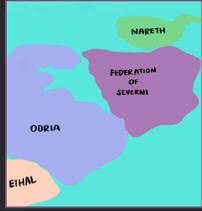
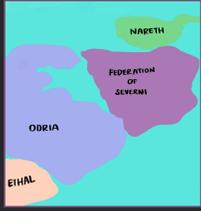
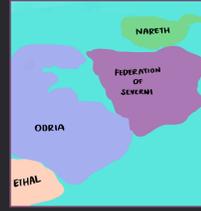
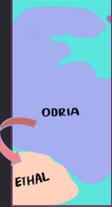
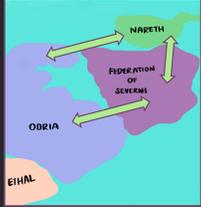
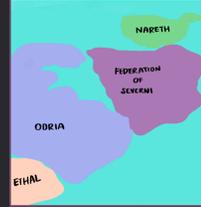
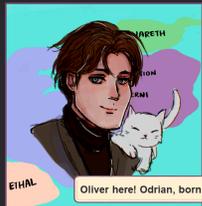

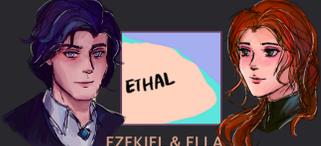

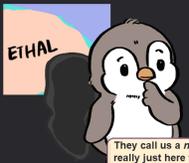
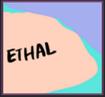
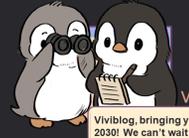
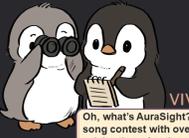
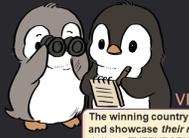
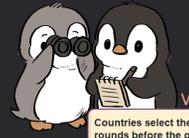
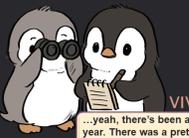
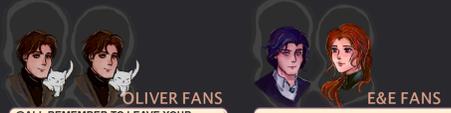
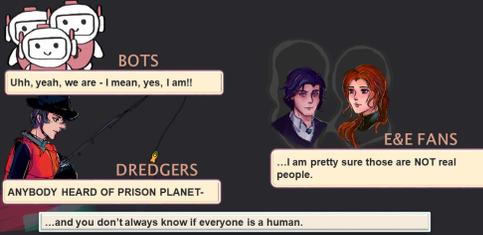

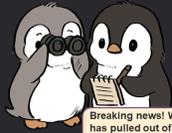 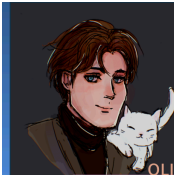 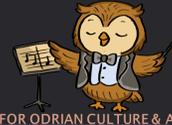

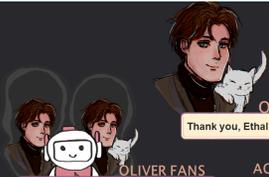 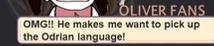 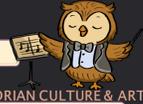 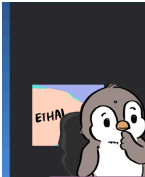 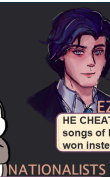 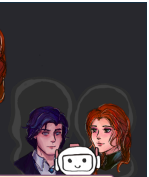

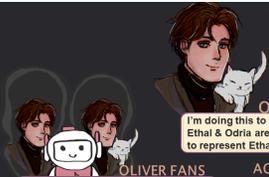 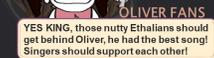 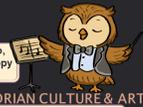 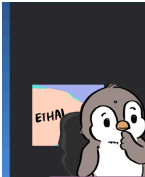 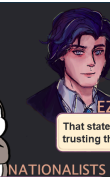 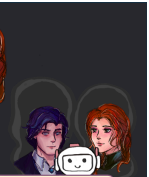